\documentstyle[12pt,aasms4]{article}

\slugcomment{Submitted to \apjl}
\lefthead{de Bernardis et al.}
\righthead{CMB anisotropy at degree angular scales ...}
\begin{document}

\title{ CMB anisotropy at degree angular scales 
and the thermal history of the Universe  }

\author{Paolo de Bernardis}
\affil{Dipartimento di Fisica, Universit\`a di Roma La Sapienza}
\centerline {\it P.le A. Moro 2, 00185 Roma, Italy. }
\centerline {\it e-mail: debernardis@roma1.infn.it}

\author{ Amedeo Balbi, Giancarlo De Gasperis}
\affil{Dipartimento di Fisica, Universit\'a di Roma Tor Vergata}
\centerline {\it Via della Ricerca Scientifica, 00133 Roma, Italy }
\centerline {\it e-mail: balbi@roma2.infn.it, degasperis@roma2.infn.it, }

\author{Alessandro Melchiorri}
\affil{Dipartimento di Fisica, Universit\`a di Roma La Sapienza}
\centerline {\it P.le A. Moro 2, 00185 Roma, Italy. }
\centerline {\it e-mail: melchiorria@roma1.infn.it}

\author{Nicola Vittorio}
\affil{Dipartimento di Fisica, Universit\'a di Roma Tor Vergata}
\centerline {\it Via della Ricerca Scientifica, 00133 Roma, Italy }
\centerline {\it e-mail: vittorio@roma2.infn.it}

\begin{abstract}
 We study the anisotropy of the cosmic microwave background (CMB) in
cold and mixed dark matter (CDM and MDM) models, with non scale-invariant primordial power spectra
(i.e. $n\neq 1$) and a late, sudden reionization of the intergalactic medium
at redshift $z_{rh}$. We test these models against recent detections of CMB
anisotropy at large and intermediate angular scales.
We find that current CMB anisotropy measurements cannot discriminate
between CDM and MDM models. Our likelihood analysis indicates that 
models with blue power spectra ($n\simeq 1.2$) and a reionization at
$z_{rh}\sim 20$ are most consistent with the anisotropy data considered
here. Without reionization our analysis gives $1.0 \le n \le 1.26 $ (95$\%$ C.L.) for
 $\Omega_b = 0.05$. 
\end{abstract}


\section{Introduction}

The COBE/DMR experiment (\cite{smo}) has revolutionized the field of
structure formation, providing the first robust evidence for primary, large
scale anisotropy of the cosmic microwave background (CMB). On these angular
scales CMB anisotropies are expected because of potential fluctuations at
last scattering (\cite{sw}). These linear fluctuations are
predicted to be constant in time, and then provide a unique tool for
determining the initial conditions out of which large scale structure formed.
Because of the low COBE/DMR angular resolution and of the correspondingly 
large
cosmic variance (\cite{abb}), these initial conditions can not be yet
determined with high precision (\cite{sca}, \cite{ben94}). In spite of this,
scale-invariant initial conditions, a robust
prediction of inflation, are indeed consistent with 
the COBE/DMR four years data
(\cite{ben96}). It is worth remembering that the analysis of these data,
based on the assumption of a flat universe, is completely insensitive to the
chemistry of the universe. It seems difficult to reconcile the bulk of the
observational data (CMB anisotropy, bulk flows, galaxy correlation function,
etc.) with a model which is not dynamically dominated by cold dark matter
(CDM). However, it does not seems probable that CDM can provide the closure
density: if this were the case, the universe would probably be more
inhomogeneous than observed on small scales. The excess power on these scales
can be reduced, for example, by considering a low density ($\Omega_0\sim 0.2$)
CDM dominated model, which is flat because of a 
suitable cosmological constant, 
or a mixture of cold and hot dark matter, i.e. mixed dark
matter (MDM) models. In this paper we will focus on the latter scenario.
However, just because of the power reduction on small scales, in a MDM model
structure formation tends to be a too recent process. This difficulty can be
alleviated by considering initial conditions that are not exactly scale
invariant. For example, it has been shown that the abundance of rich cluster
in MDM models can be better reconciled with the X-ray luminosity function by
assuming initial ``blue'' power spectra, i.e. $P(k)=Ak^n$ with $n \ge 1$
(\cite{luc}). Blue
power spectra could overproduce small scale CMB anisotropy. However, a
possible late reheating of the intergalactic medium could have controlled the
level of anisotropy at or below the degree angular scale. If and when the
universe underwent through a phase of early reionization of the intergalactic
medium is an interesting and still open question, although from the Gunn and
Peterson (\cite{gup}) test we know that the universe must 
have been highly reionized
at redshift $z\le 5$, and the presence of heavy elements in the intracluster
gas suggest that a considerable energy release occurred during the earliest
stages of galaxy formation and evolution. The recent detections of cosmic
microwave background (CMB) anisotropy at degree angular scales provide a new
and powerful way for investigating this issue. In fact, the level of detected
anisotropies at these scales can in principle discriminate among
different assumptions for the thermal history of the universe. Also, using
large and intermediate angular scales measurements increases the lever arm for
determining the primordial spectral index. A definitive answer to these and
other open questions will come when space missions (e.g. COBRAS/SAMBA and
 MAP) will provide a robust and definitive picture of the
intermediate angular scale anisotropy. Meanwhile, the number of experiments
reporting detections of anisotropy at degree angular scales has 
increased up to a couple of tens (see Table I below). 
Because of their sub-degree angular
resolution, these experiments are quite effective in testing different
reionization histories. So, the goal of this paper is to discuss if these
anisotropy data, together with COBE/DMR, discriminate among different
assumptions for the primordial spectral index $n$ and for the thermal history
of the universe.

\section{Theoretical calculations}
We consider a flat universe ($\Omega_0=1$) composed by baryons ($0.03\le
\Omega_b\le 0.07$), cold dark matter ($\Omega_{CDM} = 1- \Omega_b -
\Omega_\nu$), one family of massive neutrinos ($\Omega_\nu = 0.3$), 
photons and
two families of massless neutrinos. For age considerations we fix
$H_0 = {50  km~{s}^{-1} / Mpc}$, i.e. $h = 0.5$. 
The basic equations for describing the
time evolution of density fluctuations in these different cosmic components
have been already derived (\cite{peb3}, \cite{peb1}, \cite{peb2}). 
In Fourier space, they are:

$$ \dot {\cal G} + i {k\mu\over a} {p\over E}{\cal G} =
{1\over 4} \Phi \eqno(1)
$$
$$
\dot I_\nu + i{k\mu \over a}I_\nu = \Phi
\eqno(2)
$$
$$
\pmatrix{\dot I_\gamma \cr \dot Q_\gamma \cr} +
i{k\mu \over a} \pmatrix{I_\gamma \cr Q_\gamma \cr} =
\Phi \pmatrix{1 \cr 0 \cr} - \sigma_{\scriptscriptstyle T}n_e\left\{ {\cal V}
+4i\mu v \pmatrix{1 \cr 0\cr} - \pmatrix{I_\gamma \cr\ Q_\gamma \cr} \right\}
\eqno(3)
$$
$$
{\cal V} = {3\over 16}
{\int_{-1}^1{
\pmatrix{3 - {\mu'}^2 -\mu^2 + 3\mu^2{\mu'}^2 & 1-{\mu'}^2
- 3\mu^2\left(1-{\mu'}^2\right) \cr
1-3{\mu'}^2 - \mu^2 + 3\mu^2{\mu'}^2 & 3-3{\mu'}^2 - 3\mu^2
\left(1-{\mu'}^2\right)\cr} \pmatrix{I_\gamma \cr\ Q_\gamma \cr}
d{\mu'} } }
$$
$$
\ddot h + 2{\dot a\over a}\dot h = 8\pi G
\left( \rho_{\scriptscriptstyle B}\delta_{\scriptscriptstyle B}
+\rho_{\scriptscriptstyle CDM}\delta_{\scriptscriptstyle
CDM} + 2 \rho_\gamma\delta_\gamma + 2\rho_\nu\delta_\nu
+ 2 \rho_{\nu m} \Delta_{\nu m}\right) \eqno(4)
$$
$$ \dot h_{33} - \dot h = {16 \pi G a\over k} \left(
\rho_{\scriptscriptstyle B}v + \rho_{\nu m} f_{\nu m} +
\rho_\gamma f_\gamma + \rho_\nu f_\nu \right) \eqno(5)
$$
$$
\dot \delta_{\scriptscriptstyle B} =
{\dot h\over 2} - i {k v\over a}
\eqno(6)
$$
$$
\dot v + H(t) v =
\sigma_{\scriptscriptstyle T} n_e \left( f_\gamma -
{4 \over 3} v \right)
\eqno(7)
$$
$$
\dot \delta_{\scriptscriptstyle CDM} =
{\dot h \over 2}
\eqno(8)
$$
where $\Phi=\left( 1-\mu^2 \right) \dot h - \left(1 - 3\mu^2\right)\dot
h_{33}$. Eq.(1) describes the time evolution of the (Fourier transformed)
fluctuation in the phase space distribution of massive neutrinos: $ {\delta
{\cal F}} \equiv Y { \cal G}$, where $Y= ye^y(e^y+1)^{-2}$, $y\equiv p/T_\nu$
is the ratio between the neutrino momentum $p=\sqrt{E^2-m_\nu}$ and the
neutrino temperature (we use natural units), proportional to the CMB
temperature, $T_\nu=(4/11)^{1/3} T_\gamma$. The massive neutrino background
density is then $ \rho_{\nu m}(t) = N_{\nu m} \pi^{-2} T_\nu^4(t) \psi^{-1} $
where $\psi = \int_0^\infty dy\, y^2 \epsilon \left( e^y+1 \right)^{-1}$,
$N_{\nu m}$ is the number of neutrino families with a non vanishing rest mass
($N_{\nu m} = 1$ in our case), $\epsilon = E/T_\nu$, and each flavour state has
two helicity states. We expand ${\cal G}$ in Legendre polynomials, 
${\cal G} =
\sum_{\ell=0}^{40} g_\ell P_\ell(\mu)$, and transform Eq.(1) in a set of
coupled differential equations for the multipoles $g_\ell$ (
see e.g. \cite{bono}). The massive neutrino density contrast reads
$ \delta_{ \nu m}(k,t) = \psi^{-1} \int dy \, y^2 \epsilon Y g_0$
while
$
\Delta_{\nu m} = \psi^{-1} \int dy\, y^2 [y^2
+1/2 (m_\nu/T_\nu )^2 ] \epsilon^{-1}Y g_0.
$
For $\epsilon \simeq y $ and $\epsilon \simeq m_\nu $, we have
$\Delta_{\nu m} \simeq
\delta_{\nu m}$ and $\Delta_{\nu
m}=\delta_{\nu m}/2$, respectively. Finally, $f_{\nu m }
= \psi^{-1} \int dy y^3 Y g_1 $.
The integrals in $y$-space giving $\rho_{\nu m}$, $\Delta_{\nu m}$,
$\delta_{\nu m}$ and $f_{\nu m}$ are performed numerically with a
16 points Gauss-Laguerre integration method.

Eq.(2) describes massless neutrinos, and 
it is obtained by integrating in
$y$- space Eq. (1) in the ultra-relativistic limit 
($\epsilon\simeq y$): $I_\nu
= 4{\cal G}$. Again we expand $I_\nu$ in Legendre polynomials,
$I_\nu=\sum_{\ell=0}^{40} s_\ell P_\ell (\mu)$ 
and transform Eq. (2) in a set of 40
coupled differential equations. 
The density fluctuations of massless neutrinos
is $\delta_\nu = s_0 $, while $f_\nu = s_1/3 $. 
We numerically follow the evolution of fluctuations in this
component only when the perturbation proper wavelength is larger than one
tenth of the horizon. 
Afterwards, free streaming rapidly damps fluctuations in this hot component.

Eq.(3) describes fluctuations in the $I$ and $Q$ Stokes parameters of the 
CMB (see e.g. \cite{mel}) 
and differs from Eq. (2) because of the
collisional term, proportional to the Thomson cross section,
$\sigma_{\scriptscriptstyle T}$, and to the electronic number density, $n_e$.
The latter is evaluated assuming a standard recombination history
(\cite{jon}), with a helium mass fraction $Y_{He} = 0.23$.
We expand $I_\gamma$ and $Q_\gamma$ in Legendre polinomials:
$I_\gamma=\sum_l \sigma_l P_l$ and $Q_\gamma=\sum_l q_l P_l$.
The number of harmonics is automatically increased up to $ \ell \le 5000$
for $k \le 1 Mpc^{-1}$.

The correlation function (acf) of the temperature fluctuations can be
written as
$$ C(\alpha,\theta_B) = {1\over 4\pi}\sum_{\ell=2}^{\infty}
(2\ell+1) C_\ell P_\ell(\cos\alpha)
\exp[-(\ell+1/2)^2\theta_B^2]
\eqno(9)
$$
where $\theta_B$ is the dispersion of a Gaussian
approximating the angular response of the beam, and
$$
C_\ell = {{\cal A}^2\over 32\pi^2} {A_{COBE}\over(2\ell +1)^2}
\int {\rm d}k ~k^{2+n} |\sigma_\ell(|{\vec k}|)|^2,
\eqno(10).
$$
We define the parameter ${\cal A}^2 \equiv A/A_{COBE}$ as 
the amplitude $A$ of the
power spectrum (considered as a free parameter) in units of $A_{COBE}$, the
amplitude needed to reproduce
$ C(0^o,4^o.24) = (29 \mu K)^2 $, as observed by COBE-DMR (\cite{ben96}).

Performing the integral in Eq.(10) with high accuracy at high $\ell$'s 
requires
a very good sampling of the $\sigma_\ell$'s in $k$-space, and this is a heavy
computational task. To avoid this problem we sample the interval
$-5<log_{10}k<0$ with a step $\Delta log_{10}k=0.01$, 
we evaluate the integral
in Eq.(10), and we use a smoothing algorithm to suppress the high frequency,
sample noise. The $C_\ell$'s obtained in this way differ from those obtained
with a much denser $k$-space sampling ($\Delta log_{10}k=0.001$) by only a
fraction of percent up to $\ell \le 2000$.

The time evolution of the baryon and CDM density contrasts and of the baryon
peculiar velocity are described by Eq.(6), (8) and (7) respectively. 
The system
is closed by Eq.(4) and (5) describing the field equations for the trace and
the $3-3$ component of the metric perturbation tensor.

 We numerically integrate the previous equations from redshift $z=10^7$ up to
the present. In the following we will also assume that the universe reionized
instantaneously at redshift $z_{rh}<<1000$, and remained completely reionized
up to the present. Because of the low baryonic abundance, we need reionization
at high redshifts in order to substantially suppress the anisotropy at degree
scales. For $\Omega_b=0.05$ and $z_{rh}\le 70$, there is a probability lower
than 50 $\%$ for a photon to scatter against a free electron at $z < z_{rh}$.

\section{Data analysis}
We have selected a set of 20 different anisotropy detections obtained by
different experiments, or by the same experiment with different window
functions and/or at different frequencies. For each detection, labeled by the
index $j$, we report in Table 1 the detected mean square anisotropy, $\Delta
_j^{(exp)}$, and the corresponding 1--$\sigma$ error, $\Sigma_j^{(exp)}$. In our
error estimates, the calibration error was added in quadrature to the
statistical error. When not explicitly given, we estimated the mean square
anisotropy as follows: $\Delta_j^{(exp)}= Var[\{\Delta T_i\}_j] - \sigma_j^2$,
where $\{\Delta T_{i}\}_{j}$ are the published anisotropy data of the $j$-th
experiment, $Var[\{\Delta T_i\}_j]$ is the variance of the data points, and
$\sigma_j^2$ is the mean square value of the instrumental noise.

Theoretically, the mean (over the ensemble) squared anisotropy is given by a
weighted sum of the $C_\ell$'s:
$$
[\Delta_j^{(th)}] =
{1 \over 4\pi} \sum_\ell (2 \ell +1) C_\ell W_{\ell,j}=
  \overline{ {C_{\ell_{eff}}}\over{4\pi}} \sum_\ell 
(2 \ell +1) W_{\ell,j}
\eqno (13)
$$
where the windows function $W_{\ell,j}$ contains all the experimental details
(chop, modulation, beam ,etc.), and $\overline{ C_{\ell_{eff}}}$ is the mean
value of the $C_\ell$'s over the window function. The effective multipole
number $\ell_{eff,j}$ is defined as follows:
$$\ell_{eff, j} =
\sum \ell (2 \ell +1) C_\ell W_{\ell, j} / \sum (2 \ell +1) C_\ell W_{\ell, j}
$$
and is listed in Table I for a scale invariant model without reionization.
Although in principle model dependent, the values of $\ell_{eff ,j}$ are
quite stable because of the narrowness in $\ell$ space of the window functions.

Using numerical simulations, which take into account scan
strategy and experimental noise, we verify that the expected
distribution for $\Delta^{(exp)}_j$ is well approximated by a gaussian, with
mean $\Delta^{(th)}_j$ and a cosmic/sampling variance
$$
\Sigma^{(th)}_j = {1 \over f_j}{1\over {8 \pi^2}} {\sum_\ell}
{(2 \ell + 1) W_{\ell,j}^2 C_\ell^2   }
\eqno (14)
$$
Here $f_j$ represents the fraction of the sky sampled by each experiment and
it is also listed in Table I.

Given the Gaussian distribution of $\Delta_j^{(exp)}$, we compute the
likelihood of the 20 (assumed independent) CMB anisotropy detections
as follows:
$$
{\cal L} ({\cal A}, n, z_{rh}) =
\prod_j {1 \over \sqrt { 2 \pi [ {\cal A}^4 (\Delta^{(th)}_j)^2 +
 \Sigma_j ^2 ]}}
\exp
\bigl\{- {1 \over 2}
 { \bigl[\Delta _j^{(exp)} - {\cal A}^2 \Delta^{(th)}_j\bigl]^2
\over
 {\cal A}^4 \bigl( \Delta^{(th)}_j \bigr)^2 +
 \Sigma_j ^2 }
\bigr\}
\eqno (15)
$$
As already stated, this is a function of three parameters:
the amplitude ${\cal A}$, the spectral
index $n$ and the reionization redshift $z_{rh}$.

For each pair $n-z_{rh}$ we select the value ${\cal A}_{max}$ which maximizes the
Likelihood (isolevels of ${\cal A}_{max}$ are shown in Fig.1 for
$\Omega_b=0.03,0.05,0.07$, respectively). The results of our theoretical
calculation are shown in Fig.2a, where we plot the $C_\ell$'s for few models,
each of them normalized with its own ${\cal A}_{max}$. 
Our results show that the
difference between a pure CDM and a MDM ($\Omega_{\nu} = 0.3$) models,
normalized with the same ${\cal A}_{max}$ is very tiny, 
$\le 2\%$ up to $\ell
\le 300$ and $\le 8\%$ up to $\ell \le 800$ 
(\cite{deg}, \cite{gat}, \cite{ma}).
So we will make
hereafter the assumption that the anisotropy pattern does not depend on
$\Omega_\nu$, at least for $\Omega_\nu\le 0.3$. This generalizes our
assumption that the CMB anisotropy depends only upon three independent
variables: ${\cal A}$, $n$ and $z_{rh}$. 
For the models shown in Fig.1a we also plot
the band-power estimate for the $j$-th experiment 
$\overline{ C_{\ell_{eff}}} = {4\pi}\Delta_j^{(exp)}/ \sum {(2\ell+1) 
W_{\ell, j}}$ 
and in Fig.1b the corresponding window functions, $W_{\ell, j}$.

The 2-D, conditional distribution $L(n,z_r|A_{max}) \equiv
{\cal L}({\cal A}_{max},n,z_{rh})$ 
has a quite distinctive peak at $n = 1.24$ and
$z_{rh} = 20$. A $P$-confidence level contour in the $n-z_{rh}$ is
obtained cutting the $L $ distribution with the isolevel $L_P$, and by
requiring that the volume below the surface inside $L_P$ is a fraction $P$ of
the total volume. In Fig.3 we plot the $L_{68}$ and the $L_{95}$ contours,
again for $\Omega_b=0.03,0.05$ and $0.07$, respectively (we actually sampled a
region of the $n-z_{rh}$ plane much larger than shown in Figure).

It is clear that the considered data set identifies a preferred 
region of the 
$n - z_{rh}$ space. The shape of the confidence contours can be
understood by noting that increasing $n$ would overproduce anisotropies:
a corresponding increase in $z_{rh}$ is thus required to damp the 
fluctuations to a level compatible with observations.  On the other hand,
if $\Omega_b$ is increased, the optical depth increases, thus requiring
a lower reionization redshift to produce the detected level of degree-scale
anisotropy. This effect dominates over the smaller increase of primary CMB
anisotropies.

The simple analysis carried out here does not take into account
details on the scan pattern and/or the beam profile.
Moreover, we know that the published error bars could not account 
properly for
correlated errors or other more subtle effects. In order to test the
significance of our analysis and its robustness against 
the published estimates
for the experimental errors, we take a drastic point of view: we consider the
best fit model ($n=1.24$, $z_{rh}=20$) normalized to COBE/DMR
(i.e. we fix ${\cal A}=1$).
For each degree--scale data point we compute the difference 
between the measured
value and the value expected from the theory.
We then associate to each data point a 1--$\sigma$ error bar equal to
the root mean square of these 19 differences (instead of the published error
bars). The corresponding results of the
likelihood analysis are in very good agreement with what is found using the
appropriate errors: $n>1$ is slightly better, especially if the reionization
redshift is increased. So we do believe that our results are little
affected by possible correlations in the errors and/or systematics.

\section{Conclusions}

Our main conclusions, derived from Fig.3, are as follows.

1. The conditional Likelihood shows a maximum for $n = 1.24$ and $z_{rh}
= 20$, for $0.03 \le\Omega_b\le 0.07$, i.e. for baryonic abundances
consistent with primordial nucleosynthesis. We have checked the stability of
this result by applying a jack-knife analysis to the considered data set,
and we have seen that the results are reasonably stable:
the best model has always $n=1.24 (\pm 0.02)$ and 
$z_{rh} = 20 (\pm 10)$, unless COBE data are excluded 
(in that case $n=1.40$ and $z_{rh} = 20$).

2. The 95$\%$ confidence contours in 
the $n-z_{rh}$ plane include a wide range
of parameters combinations. This means that the presently available data set
is not sensitive enough to produce "precise" determinations for $n$ and
$z_{rh}$; systematic and statistical errors in the different experiments are
still significant. 

3. If we exclude an early reionization of the intergalactic medium 
($z_{rh}=0$) we get the following $95 \%$ confidence level 
estimates for the spectral
index: $1.02 \le n \le 1.28$ ($\Omega_b = 0.03$); 
$1.00 \le n \le 1.26$ ($\Omega_b= 0.05$); 
$0.96 \le n \le 1.24$ ($\Omega_b = 0.07$). This has to be compared to
the results from COBE-DMR alone: $n=(1.3 \pm 0.3)$, at the $68 \%$
confidence level ( \cite{ben96} ). So, in spite of their still low signal
to noise ratio, the degree scale experiments already allow to better 
constrain the spectral index, although still at the $10\%$ level.
Note that the "standard", flat model with no reionization is close 
but not always inside the $95 \%$ confidence contour.
However, this result is determined mainly by the Saskatoon experiment.
In fact, we tested the stability of our results 
by repeating the analysis several times, excluding 
one experiment at the time. If we do not consider reionization,
the $95\%$ C.L. interval $1.00 \le n \le 1.26$ does not change more than a 
few $\%$ excluding either Argo, CAT, MAX, MSAM, South Pole or Tenerife. 
The lower limit drops to $0.86 \le n$ if the Saskatoon data are excluded, 
while the upper limit raises to $n \le 1.76$ if the COBE data are excluded. 
This test also shows that neglecting the correlation due to overlapping
sky coverage (e.g. Tenerife and COBE, and/or MSAM and Saskatoon) does not
change significantly the results of our analysis. 

4. For scale invariant models with no reionization, the height of the first
"Doppler" peak occurs at $\ell \simeq 220$ and is a factor of 
$\simeq 5.6$ higher than
the Sachs-Wolfe plateau at low $\ell$'s. For $n=1.2$ the peak amplitude is
roughly a factor $1.5$ higher than in the scale-invariant case. A complete
reionization from $z_{rh} \simeq 20$ up to the present suppress 
the peak by roughly $20\%$. Altogether, a model with $n=1.2$ and $z_{rh}=20$
has a Doppler peak a factor of $2$ higher than in the standard scale 
invariant case without reheating. 
So our analysis confirms that a Doppler peak in the
$C_\ell$ spectrum centered at $\sim 200$ is perfectly consistent with the
data.

All the models we have worked with have, as stated above, $h=0.5$. In flat
models with vanishing cosmological constant we can expect $0.4 \le h \le 
0.65$ from the estimated age of globular clusters 
(\cite{kol}). Varying $h$ modifies the amplitude and position of the
acoustic peaks in the radiation power spectrum. However, we have
verified that, in the region of the spectrum probed by current anisotropy
experiments, $20\%$ variations of the Hubble constant yeld modifications
in the spectrum very similar to those obtained varying $\Omega_b$
between 0.03 and 0.07. Also, we did not consider tensor modes in our
analysis, as they are expected to be of negligible amplitude for $n>1$
(\cite{ste}). We will address this issue in a forthcoming paper.

\newpage

\figcaption[fig1.eps]{Isolevels of the normalization amplitude
${\cal A}_{max}$, which maximizes the combined likelihood 
${\cal L} ({\cal A}, n, z_{rh})$. \label{Fig.1}}

\figcaption[fig2.eps] {Power spectra of CMB anisotropies for
different combinations of spectral index and reionization redshift 
(panel a). The data points are derived from the 
experiments listed in Table 1. In panel b we plot the corresponding filter 
functions. \label{Fig.2}}

\figcaption[fig3.eps]{ Confidence level (68 and 95$\%$) regions for
the spectral index $n$ and the reionization redshift $z_{rh}$. 
The black square marks the position of the model featuring the 
maximum likelihood. \label{Fig.3}}

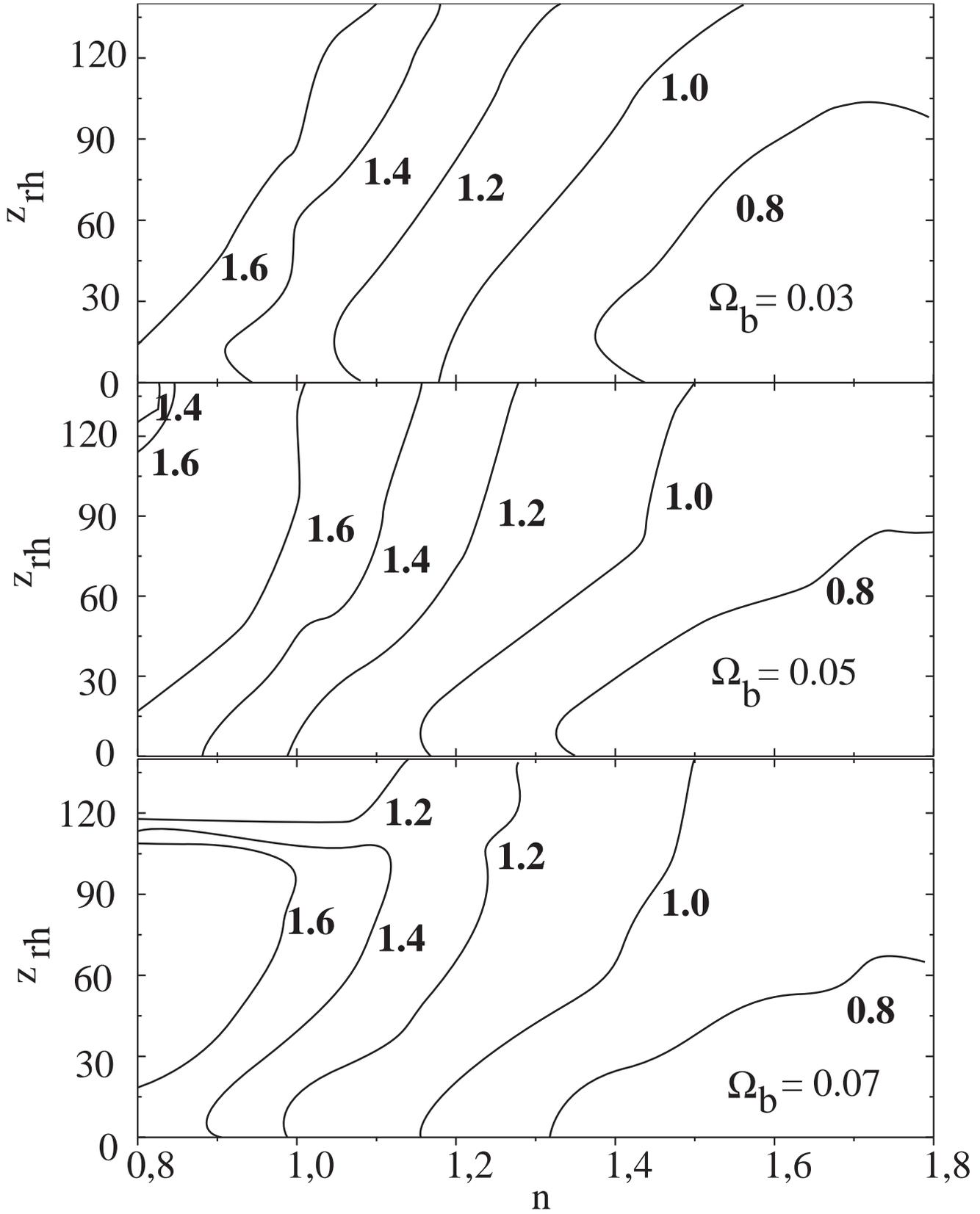
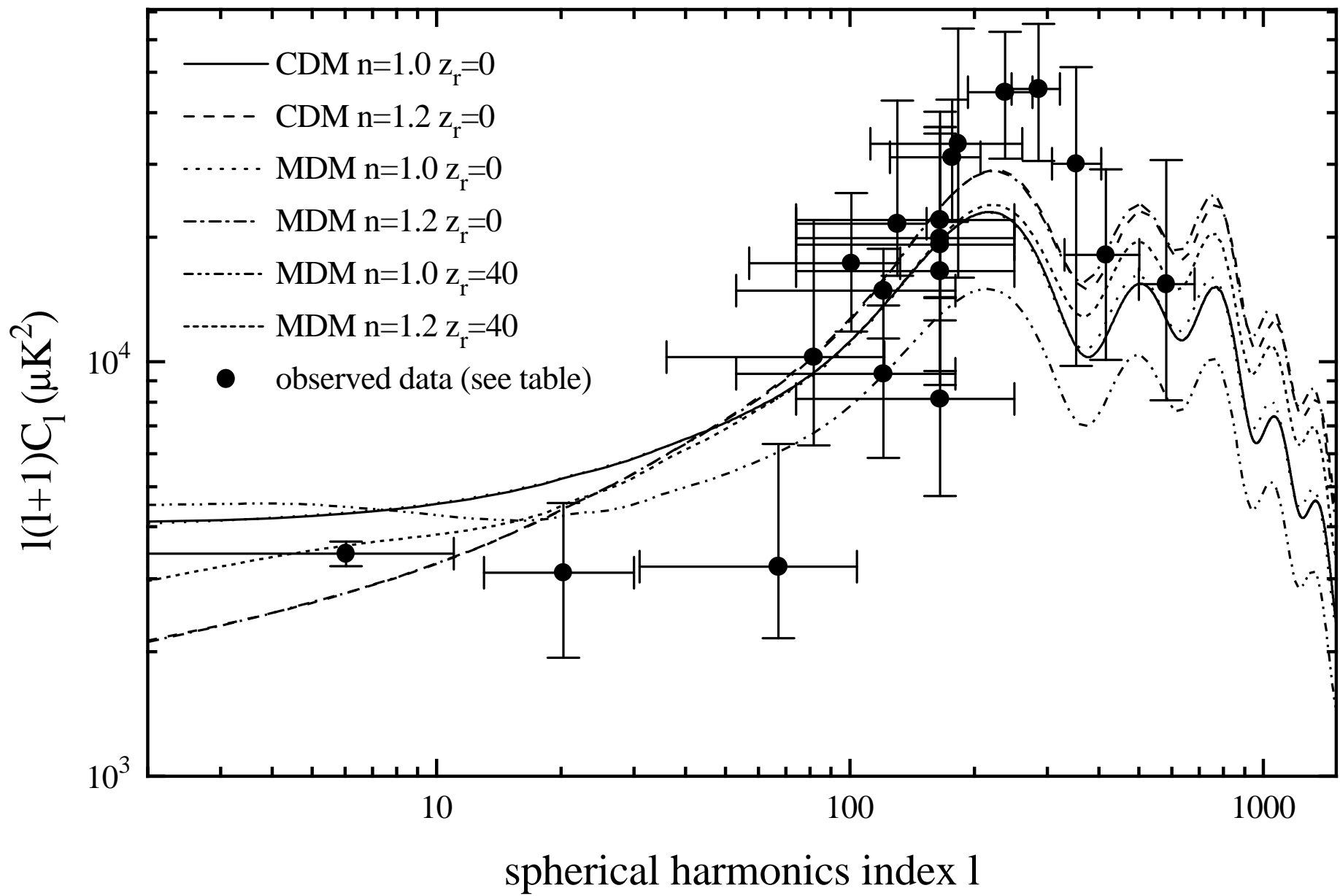
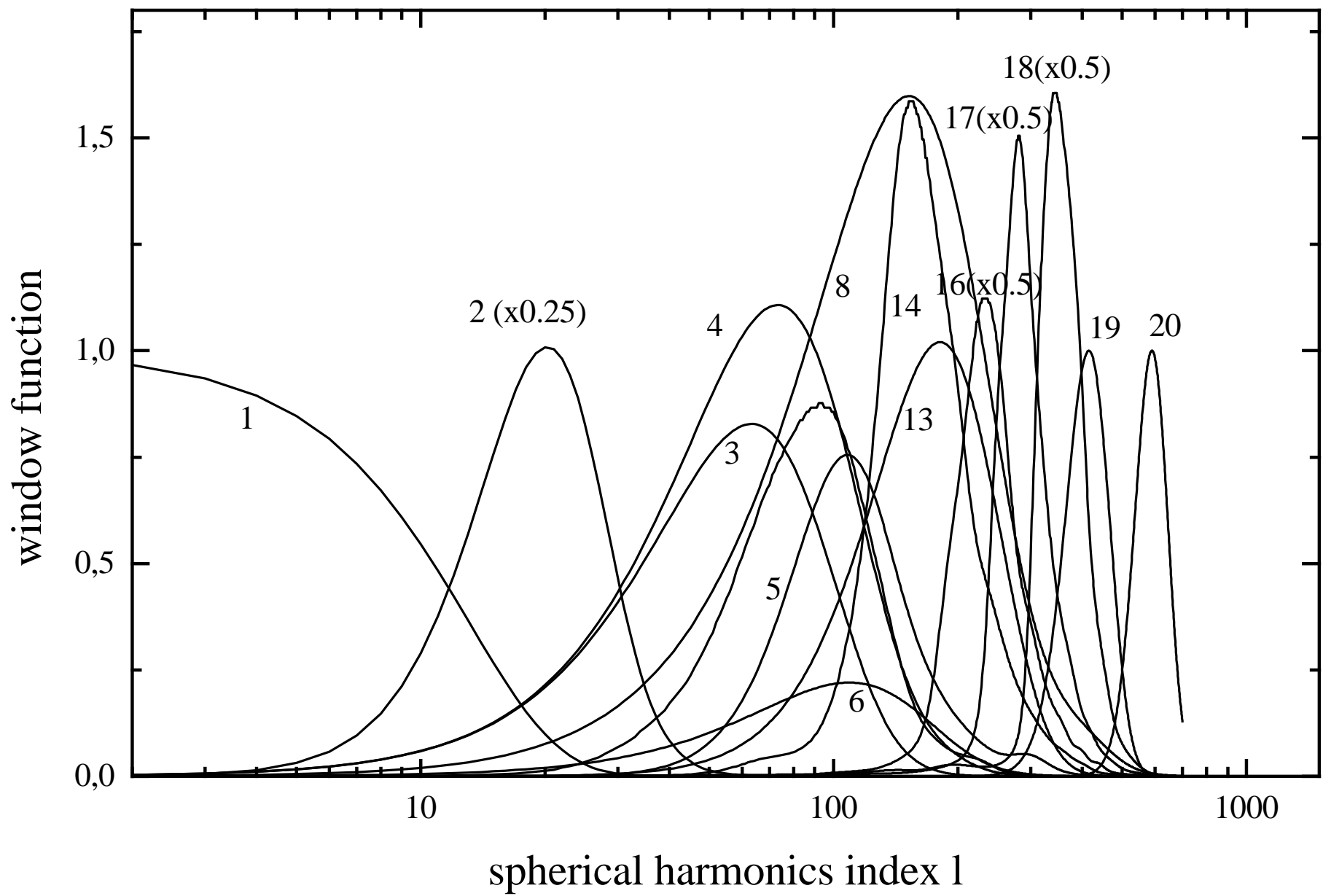
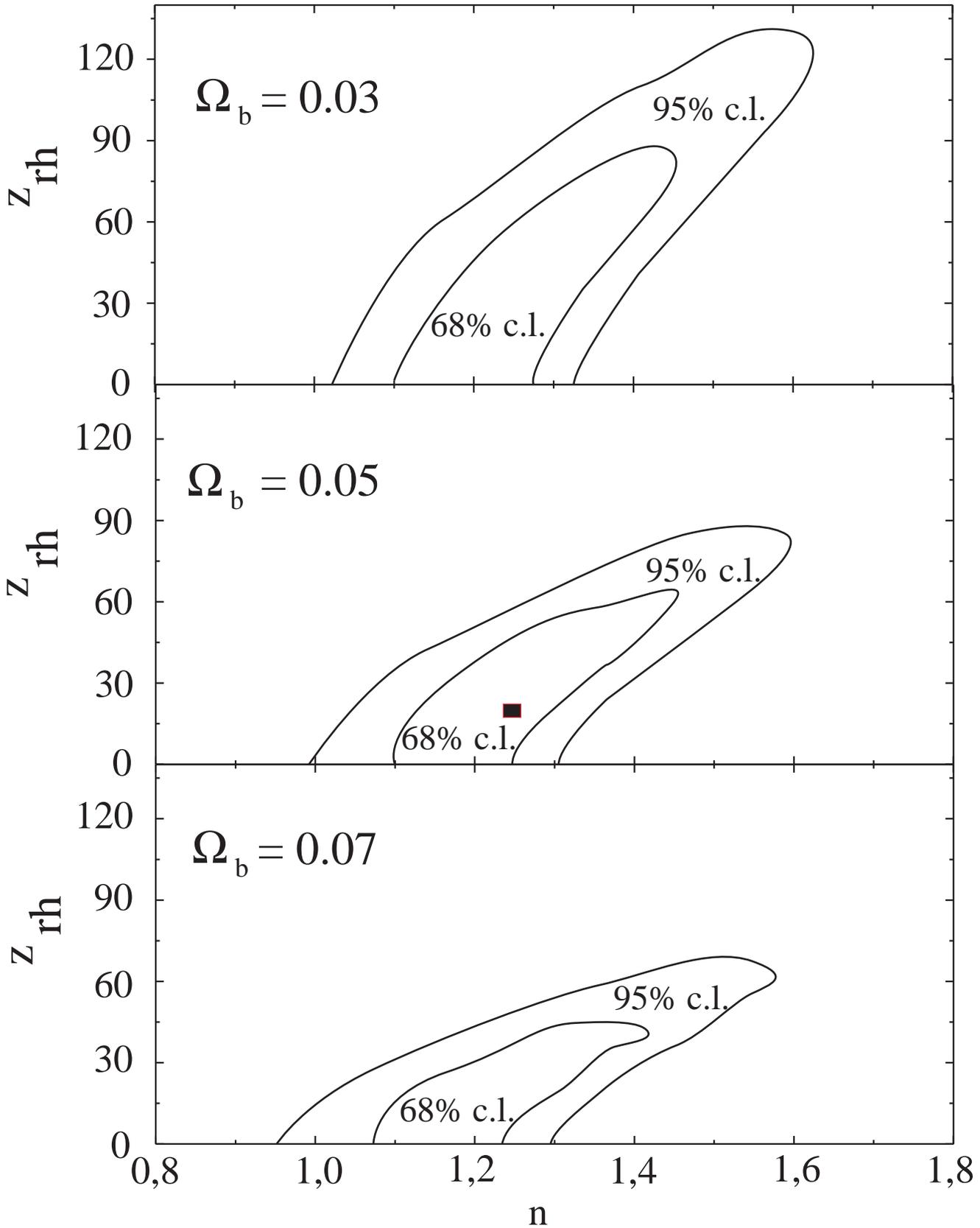
\end{document}